\pdfoutput=1
\documentclass{article}
\usepackage[a4paper, margin=1in]{geometry}
\usepackage{graphicx} 
\usepackage{cite}
\usepackage{url}
\usepackage[utf8]{inputenc}
\usepackage{amssymb}
\DeclareUnicodeCharacter{211D}{\mathbb{R}}
\usepackage{amsmath,amssymb,amsfonts}
\usepackage{subcaption}
\usepackage{caption}
\usepackage{float}
\usepackage{textcomp}
\usepackage{longtable}
\usepackage{booktabs}
\usepackage{physics}
\usepackage[table, xcdraw]{xcolor}
\usepackage{array,ragged2e}
\usepackage{lscape}
\usepackage{tabularray}
\usepackage[utf8]{inputenc}
\usepackage{amssymb}
\usepackage{amsmath,amssymb,amsfonts}
\usepackage{graphicx}
\usepackage{epsfig}
\usepackage{authblk}
\usepackage{orcidlink}
\usepackage[algo2e]{algorithm2e}
\usepackage{algorithm}
\usepackage{algpseudocode}
\title{Quantum Convolutional Neural Networks for Groundwater Heat Plume Prediction: A Surrogate Modeling Approach}
\author[1*]{\textbf{Danyal Maheshwari}~\orcidlink{0000-0001-7544-7817}}
\author[1]{\textbf{Julia Pelzer}~\orcidlink{0009-0003-1727-9626}}
\author[1]{\textbf{Miriam Schulte}~\orcidlink{0000-0003-4594-6142}}

\affil[1]{%
Simulation of Large Systems, Institute for Parallel and Distributed Systems (IPVS),\\
University of Stuttgart, Stuttgart, Germany\\

\texttt{danyal.maheshwari@gmail.com}\\
\texttt{Julia.Pelzer@ipvs.uni-stuttgart.de}\\
\texttt{Miriam.Schulte@ipvs.uni-stuttgart.de}
}

\begin{document}

\maketitle

\begin{abstract}
Quantum machine learning methods are increasingly explored for modeling complex environmental systems, including groundwater heat plume dynamics. In this work, we explore a Quantum Convolutional Neural Network (QCNN) as a surrogate model for predicting temperature variations in groundwater induced by geothermal heat pumps in the city of Munich. To comply with the scalability constraints of current quantum hardware, the original high-dimensional simulation output is reduced to a compact set of representative parameters that serve as training targets for the surrogate. The proposed QCNN architecture consists of a quantum convolutional layer, a quantum pooling layer, and a fully connected quantum readout stage. Convolution and pooling operations are realized via parameterized quantum circuits based on rotational gates and measurement-driven decoding, while a Hamiltonian-inspired feature-encoding scheme is used to prepare informative input states on the quantum device. We evaluate the QCNN across multiple execution backends, including an ideal statevector simulator, a noisy simulator, IBM’s 127-qubit \textit{Kyiv} quantum processor, and the same hardware augmented with advanced error-mitigation techniques. Realistic noise models are employed to approximate device behavior and to assess the impact of mitigation strategies. Model performance is benchmarked using mean squared error (MSE) on both training and testing sets. The results show that, although classical neural networks still achieve the highest predictive accuracy, the QCNN attains competitive and consistent performance on simulators and exhibits noticeable improvement under error-mitigated hardware conditions. These findings indicate that quantum-enhanced surrogate modeling is a promising direction for future groundwater temperature prediction as quantum hardware and error-mitigation techniques continue to mature.

\end{abstract}

\section{Introduction}

Climate change poses one of the most significant challenges of the 21st century, with burning fossil fuels as one of the main contributors to greenhouse gases. The heating and cooling of private households in Germany accounts for 123.4~million tonnes of direct carbon dioxide emissions in 2020~\cite{umweltbundesamt}. This is about 17~\% of the annual emission of CO$_2$ equivalents in 2020~\cite{bundesministerium}.
To mitigate climate change, cities such as Munich, Germany, need to transition to renewable and more efficient energy sources, such as shallow groundwater heat pumps (GWHP) for heating and cooling houses~\cite{holihan98}.

According to regulations, new GWHPs must not be installed upstream of existing systems in a manner that significantly reduces their efficiency. Additionally, groundwater temperature alterations must remain within permissible limits to preserve microbial activity and fauna composition, both essential for groundwater purification~\cite{bw_leitfaden}.
This motivates us to model the temperature and velocity in the groundwater body to predict the direction and intensity of heat plumes.
A pilot project geoKW~\cite{geokw22} in Munich aims to model the whole subsurface temperature field of the region of Munich with the numerical simulation software Pflotran~\cite{pflotran} to optimize the positions of as many heat pumps as possible. This optimization requires the temperature field modeling of a lot of different heat pump constellations.

Simulations are very accurate but require long simulation times and a new run for every change in input parameters. The need for real-time applications and the modeling of a lot of scenarios calls for surrogate models with less accuracy but higher speed. 
Surrogate models split into those estimating heatplume characteristics, i.e., its length and width, mainly done by analytical models like the Linear Advective Heat transport Model (LAHM)~\cite{kinzelbach87}, and those estimating a spatially resolved temperature field. The characteristics are only feasible to fully represent the heat plume of a single heat pump in a homogeneous aquifer. In a heterogeneous aquifer, spatially resolved methods like the 1HP-CNN approach in~\cite{pelzer2024}, based on Convolutional Neural Networks (CNNs), are necessary.


The rapid advancement of quantum computing (QC) in recent years has sparked considerable interest in both theoretical and applied domains, heralding its potential to revolutionize real-world applications. The fundamental principles of quantum mechanics have necessitated a reevaluation of QC's capabilities and implications. Significant progress in physics throughout the 21st century has led to enhanced material purity and more sophisticated observation techniques, rendering previously elusive quantum phenomena increasingly detectable and measurable \cite{AdvancesQuantumComputing2023, QuantumSensors2025}. 

Quantum Machine Learning (QML) emerges as a multidisciplinary field at the confluence of QC and Machine Learning (ML), two of the most dynamic areas in contemporary computer science. It also allows for the beehive-like convergence of ideas that lead to cross-pollination methodologies from one domain to deal with problems in other areas \cite{maheshwari2022quantum}. In light of growth in the amount and complexity of data by exponential proportions, this offers really promising grounds for the accomplishment of QML tasks that may well have been out of reach by its classical counterpart in computation \cite{Fan2026BridgingQCLearning, maheshwari2023quantum}.

QC's impact on ML algorithms and methodologies is an increasing research area, where many interests come from both the academic and industrial sectors. Indeed, recent advances in the field have led to significant improvements in ML algorithms, inspired by quantum mechanical principles such as superposition and entanglement. Quantum algorithms, therefore, have the prospect to outperform their classical counterparts for specific applications, mostly concerning optimization, pattern recognition, and classification of data \cite{QMLRecentAdvances2025}.

The QML field is continually evolving in quantum algorithms, and architectures are being sought to solve machine learning problems. In fact, such work is really needed; it will enable quantum systems to pass the computational bottleneck of the current systems and help analyze new data and solve new problems that were beyond the capability of current computing systems. This combination of QC and ML not only improves existing algorithms but also opens new avenues for completely new ways of information processing and artificial intelligence \cite{RecentAdvancesQML2023, QMLRecentAdvances2025}.

One promising area is the integration of quantum circuits into neural network architectures, leading to the development of Quantum Convolutional Neural Networks (QCNNs)\cite{maheshwari2025predicting}. QCNN makes use of some of the inherent properties of quantum systems, such as superposition and entanglement, in order to extract nonlinear features and relationships present in the input data much more powerfully than previous methods.

This research is organized into five main parts: Section II, groundwater heat plumes. Following this, we examine the Heat plume dataset. Section III delves into QCNN algorithms relevant to our study. Section IV outlines our experimental design and discusses the findings in detail. The final section presents our concluding remarks and insights.

\section{Related work}
Current modeling approaches include numerical simulations, analytical models, and deep learning surrogate models.


\subsection{Analytical models}
Three models for calculating the temperature difference caused by a heat pump at a distance $(x,y)$ and time $t$ are LAHM, Planar Advective Heat transport Model (PAHM), and Radial Heat transport Model (RHM), as compared in \cite{Poph20}, \cite{Ohmer2022}.
These models assume two-dimensional, transient flow in confined aquifers with homogeneous hydro-geological parameters. The choice of model depends on the scenario: LAHM handles line sources with continuous injection and a constant background flow, PAHM is suited for planar sources with moderate flow, and RHM applies when background flow is negligible. 
Given the requirements of an injection well, LAHM is the most appropriate model.

LAHM~\cite{kinzelbach87} performs best under two conditions: the energy usage remains below 45,000kWh/year, and groundwater flow velocities exceed 1m/day. It simplifies convective and dispersive heat transport
\begin{equation}\label{eq:lahm}
\begin{split}
\Delta T (x,y,t) & = \frac{Q \Delta T_\text{inj}}{4 n_e M v_a \sqrt{\pi \alpha_T}} \cdot \exp{(\frac{x - r}{2 \alpha_L})} \cdot \frac{1}{\sqrt{r}} \cdot \text{erfc }(\frac{r - v_a t/R}{2 \sqrt{v_a \alpha_L t/R}})\\
r & = \sqrt{x^2 + y^2 \cdot \frac{\alpha_L}{\alpha_T}},
\end{split}
\end{equation}
with hydro-geological parameters of aquifer thickness $M$ (in $m$), the effective porosity $n_e$(unitless), longitudinal and transversal dispersivity $\alpha_L$, $\alpha_T$ (in $m$), the effective groundwater velocity $v_a$ (in $m/s$), and the retardation factor $R$. The operational pump parameters include the volumetric pump rate $Q$ (in $l/s$), the difference between injection temperature, and undisturbed groundwater temperature $\Delta T_\text{inj}$ (in $K$).

\begin{figure}[!htbp]
    \centering
    \includegraphics[width=\textwidth]{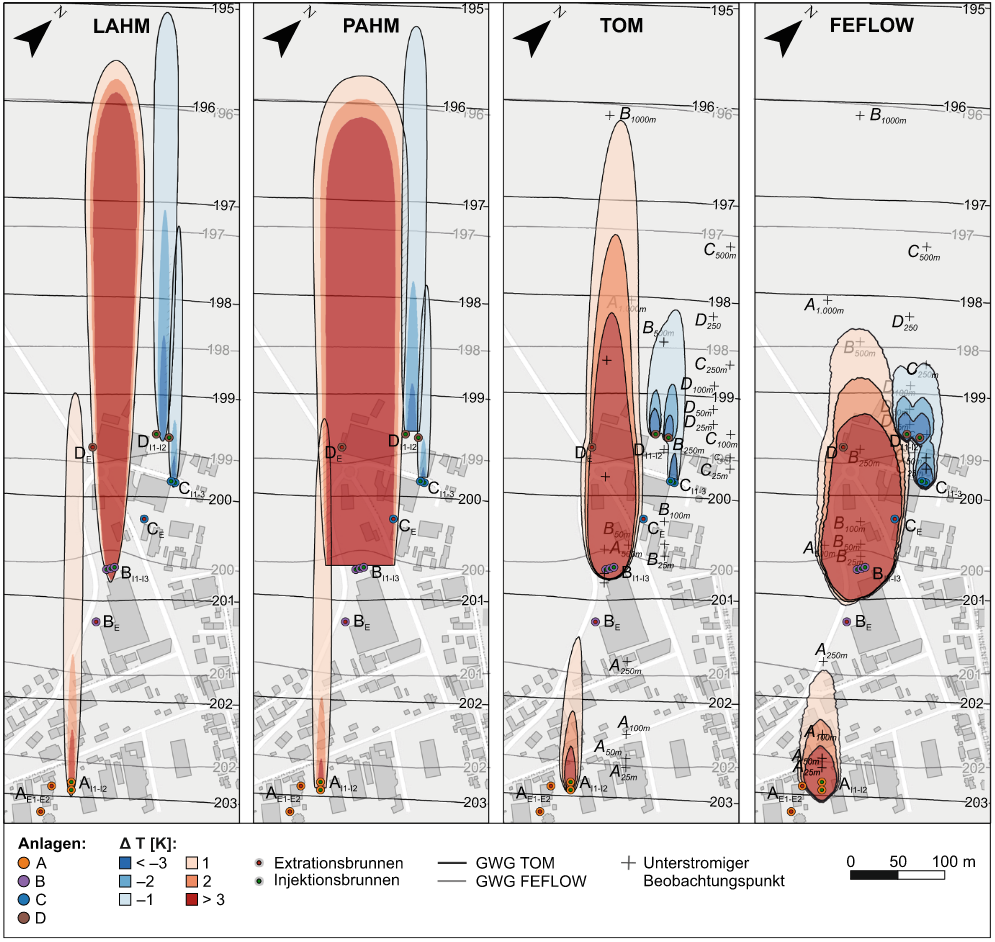}
    \caption{Qualitative, exemplary comparison between LAHM, PAHM, TOM, and FEFLOW from left to right on systems of varying size. Source \cite{Ohmer2022}.}
    \label{fig:lahm_tom_feflow}
\end{figure}

While LAHM provides quick estimates, it tends to overestimate the width of the 1K-isoline \cite{Poph20, Ohmer2022}, see Fig.\ref{fig:lahm_tom_feflow}. Additionally, it assumes confined aquifers, which is uncommon near Munich\cite{geologica}, and it can not account for near-well effects, multiple interacting plumes, or heterogeneous aquifer conditions.
Fig.~\ref{fig:lahm_tom_feflow}.

\subsection{Deep learning models}
The most successful deep learning approaches to learn spatial temperature fields from spatial subsurface parameters and heat pump locations ~\cite{davis23}, \cite{leiteritz22}, \cite{pelzer2024} are based on CNNs.

The first approach~\cite{davis23}, \cite{leiteritz22} learns the temperature field around a single heat pump from a spatially varying groundwater flow velocity field, based on synthetic datasets simulated with Pflotran~\cite{pflotran}. The chosen network architecture is a variation of TurbNet~\cite{thuerey_airfoils}, a variant of UNet~\cite{RonnebergerFB15}. UNet is an encoder-decoder CNN that includes skip connections to overcome limitations of vanishing gradients in very deep networks during backpropagation. More importantly, skip connections between early layers in the encoder part and late layers in the decoder part ensures the preservation of high-resolution features.
A limitation of the approach is that it learns from a velocity field, which has to be simulated first. The velocity fields are not contained in the measurements. 

\begin{figure}[!htbp]
    \centering
    \includegraphics[trim=1.6cm 19cm 1.8cm 2.2cm, clip]{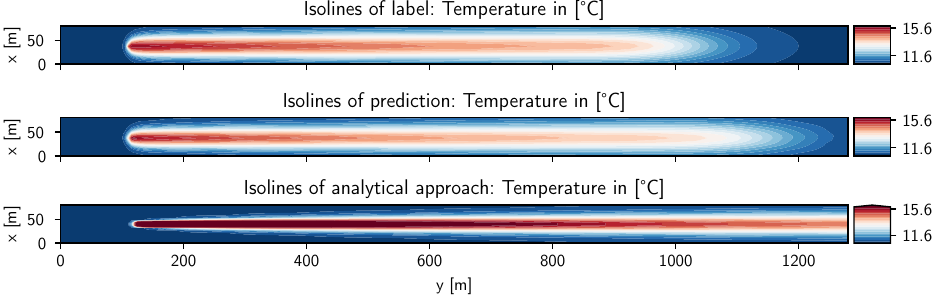}
    \caption{Comparison of 1HP-NN to analytical and simulation approaches: Isolines of the temperature field in steady-state of an exemplary data point with the simulated result on top, then the prediction from the CNN and on the bottom the calculated result from the analytical approach LAHM. Example with v < 1m/day for which LAHM is not intended. Source~\cite{pelzer2024}.}\label{fig:example_sim_nn_lahm_1hp}
\end{figure}

Another CNN-based approach of~\cite{pelzer2024} learns from directly measureable input parameters of a homogeneous permeability and pressure gradient field. The two-staged approach first estimates the heat plumes of isolated heat pumps. In a second step, these estimations are corrected according to neighboring heat plumes.
During the first step, the same box of a temperature field is estimated as the input parameter boxes. The optimized model architecture is also based on UNet, including 3$\times$3 convolutional layers in the encoder and 3 in the bottleneck, and a kernel size of 5$\times$5.
The model is trained for $\approx$~18,500~epochs on 1,000~data points, achieving an absolute error of~0.046\textdegree C averaged over 100 test data points on 16$\times$256~cells.
The cell with the highest average error is directly at the injection point with $\approx$~0.21\textdegree C. The isolines-including temperature field of an exemplary test data point estimated by LAHM and the CNN is visualized in Fig.~\ref{fig:example_sim_nn_lahm_1hp}.


Simulations bear the advantage of the highest accuracy but also the highest computation effort, which is unfeasible in real-time applications. Analytical models are the fastest but least accurate and underlie the strongest limitations. Due to their speed in application, they are currently used for real-time applications~\cite{geokw22}. Surrogate models, such as CNNs, offer a promising compromise of possibly accuracy and high speed, making surrogate models the choice for real-time applications and applications that require a lot of scenarios to be modeled.

\section{Dataset}
The models are trained on a dataset that we generated with the open source flow and transport modeling software Pflotran~\cite{pflotran}. 

The dataset consists of 100 data points, each containing one open-loop groundwater heat pump inlet well at a fixed position, centered in the x dimension and on the upstream end of the domain to leave space for the plume. 
The simulations run for 5 years and have therefore not necessarily reached a steady state. The simulated domain measures 100$\times$1280~m with a resolution of 5~m resulting in 20$\times$256~cells. 
To obtain the dataset, two properties are varied: the boundary and initial conditions of the hydraulic gradient resulting in a homogeneous pressure gradient over the domain but varying for each data point between 0.0015 and 0.0035; and a homogeneous permeability field with values between $1.03 \times 10^{-11}$ to $4.37 \times 10^{-9}$ $\text{m}^2$.
This results in a variance of flow velocities and permeability with varying width and length of the resulting heat plume from the injection well. Due to the low resolution, the maximum observed temperature depends on the subsurface velocity with which the heat is transported.

Other flow-defining parameters such as porosity or heat conductivity are constant in the domain and over the data set. 
All parameters are chosen to match the measurements from the boreholes in the Munich region according to~\cite{bottcher2022influences}, more closely described in~\cite{pelzer2024}.

To make the data set applicable for QC, we reduced the input size per data point from 2$\times$20$\times$256 (channels $\times$ width $\times$ length of the domain) to the two scalar variables of permeability and pressure gradient. The output temperature field originally contained 1$\times$20$\times$256 values. It is reduced to four parameters of length and width of the heat plume and maximum temperature and position of the maximum temperature in the flow direction. Those parameters characterize each heat plume in a homogeneous aquifer.
The length and width of the heat plume are determined by its isoline of 1~\textdegree C temperature difference to the surrounding groundwater.
The position of the maximum temperature is provided as an absolute position in the simulated box in cells.

In extreme cases of permeability and pressure gradient, the plume can extend the bounding simulation box. In those cases, the length and width are set to $NaN$. This happens in 10~cases.
Another edge case is that the flow velocity is so high that the maximum temperature difference is always below 1~\textdegree C. If so, we set the length and width to zero.

The prepared dataset with 1000 data points, including edge cases, can be found in~\cite{darus}. 
 
\section{Methodology}

\subsection{State Preparation}
The state preparation is the foundational step in preparing and processing data into QC. A typical approach in supervised learning for classification tasks is calculating a function $f$ that maps the input data $D$ to the output labels $y$, denoted as $y = f(D)$. The fundamental goal of classification is to enhance the prediction of the models.

In our cases, we have to encode data $D = (h_i, p_i)_{i=1,\ldots,M}$ with the hydraulic gradient $h_i$ and permeability $p_i$ of $M$ data points. We assume that both are given in a binary representation with $n$ bits. Any normalization of $h_i$ and $p_i$ to $[0,1]$ with $i\ \in\{\ 1,2,....,M\}$ and $M$ denotes the total number of attributes. We encode such a set of data points via basis encoding as equation (\ref{eq1}) 

\begin{equation}
    \ket{D} = \frac{1}{\sqrt M}\sum_{i=1}^{M}{\ket{(h_i,p_i)}}
    \label{eq1}
\end{equation}

In the above equation, $h$ and $p$ represent the features $n$ in the data point properties $i$, and $y_i$ is the corresponding label.

The quantum state $|\psi_i\rangle$ represents the encoded form of the classical data point $D_i$, where $|\psi_i\rangle$ belongs to the high-dimensional Hilbert space $\mathbb{C}^{2^d}$. There are several techniques to embed classical data into quantum states in higher dimensions \cite{schuld2019quantum, maheshwari2021vqc}.

\textbf{Basis Encoding:} Basis encoding is the most common technique to embed classical data into a quantum state. This method establishes a correspondence between the \(n\)-bit classical data points and the computational-basis states of the \(n\)-qubit quantum state. For example, classical 16-bit data points (0000, 0001, ..., 1111) can be encoded into the 4-qubit quantum states \(|00\rangle, |01\rangle, ..., |11\rangle\). 
To achieve a pure quantum state for data encoding, a feature map circuit is employed to project data onto computational basis states, with probability amplitudes corresponding to the sample distribution \cite{schuld2019quantum, maheshwari2021vqc}. This encoding process can be represented as:

\[
|\psi_{\mathbf{D}}\rangle = \sum_{i=0}^{2^n-1} D_i |i\rangle,
\]

where \(D_i\) are the probability amplitudes and \(|i\rangle\) are the computational basis states. If we encode the entire feature vector \(\mathbf{D}\) into the quantum state and assume that we have a total of \(n\) qubits, the overall quantum circuit for the feature map is a tensor product of individual unitary operations applied to each qubit:

\[U(\mathbf{D}) = \bigotimes_{i=1}^n U(D_i)\]

The advantage of this encoding is that you can store \(S_x\) features using only \(O(2^n)\) qubits, which constructs a tensor product of one-qubit rotations based on the data vector D.

\[
S_{\mathbf{D}} |0\rangle = \bigotimes_{i=1}^n \cos(D_i) |0\rangle + \sin(D_i) |1\rangle.
\]

\begin{algorithm}
\textbf{Input}: Training Dataset \( D = (h_i, p_i)_{i=1,\ldots,M} \), Backend \(S\), Learning Rate $\alpha$, Regularization Feature Data $F_r$, Number of Qubits $N_q$, Features $N_f$, Output Dimension $M$, Epochs $E$ \\
\textbf{Output}: Classified Output \\
\vspace{0.01cm}
\Repeat{}{
\quad \textbf{Feed Forward Propagation:}

\quad $F_{map} \leftarrow$ Data Encoding ($N_q$, $F_r$)

\quad $Q_{conv} \leftarrow$ Quantum Convolution Layer ($N_q$)

\quad $Q_{pool} \leftarrow$ Quantum Pooling Layer ($N_q$)

\quad $D_2 \leftarrow$ Linear Layers with RELU Activation ($N_q$, $2 \times N_q$, $M$)

\For{each input $D_i$ in $D$}{

\quad \textbf{Data Encoding:} Encode $D_i$ into quantum state $\left|\psi_i\right\rangle$

\quad Apply $Q_{conv}$, $Q_{pool}$ on $M_1$

\quad \textbf{Measurement:} Apply $M_2$ with RELU activation

\quad \textbf{Backpropagation:} Adam Optimizer, Loss Function: Mean Squared Error (MSE)
}}
\textbf{Return}: Predicted labels
\caption{\textbf{QCNN Algorithm}}
\label{algo}
\end{algorithm}

\begin{figure*}
    \centering
    \includegraphics[width=\linewidth]{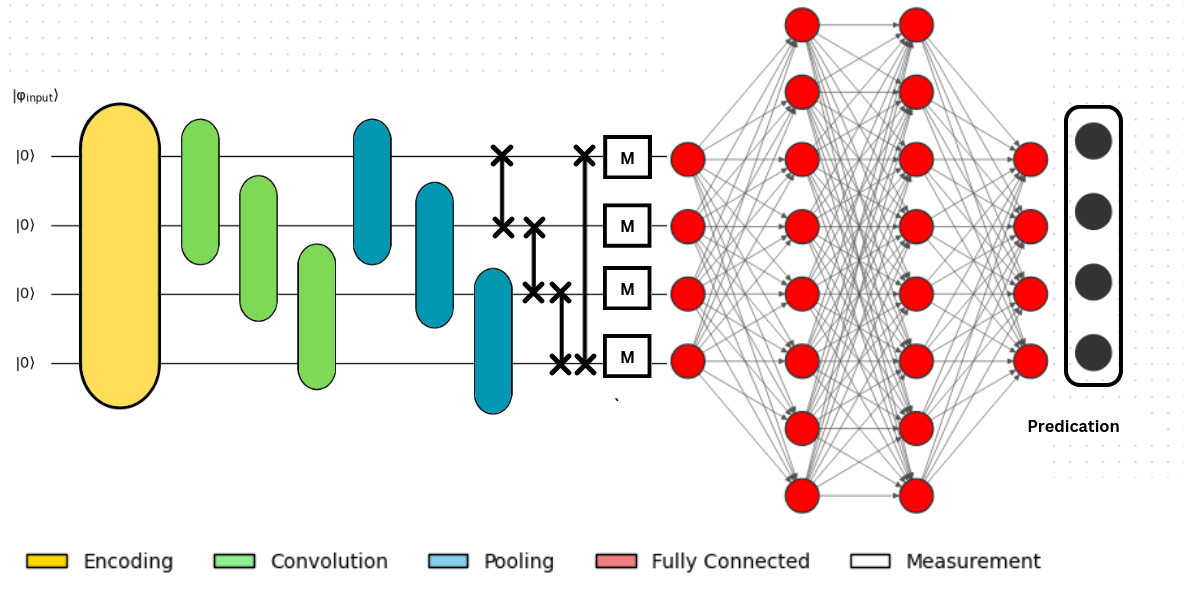}
    \caption{Schematic diagram of QCNN} 
    \label{fig:enter-label}
\end{figure*}

\subsection{Quantum Convolutional Neural Network}

Convolutional Neural Networks (CNNs) are widely used in supervised learning for tasks such as classification, regression, and pattern recognition. Quantum Convolutional Neural Networks (QCNNs) extend this paradigm to the quantum domain by exploiting superposition, entanglement, and quantum parallelism \cite{cong2019quantum, 9289439}. By combining quantum state preparation, parameterized quantum circuits, and measurement readout, QCNNs aim to enhance the representational capacity of classical CNNs, particularly for high-dimensional or strongly correlated data \cite{cong2019quantum, 9289439}.

A key component of a QCNN is the \emph{state preparation} (or quantum feature encoding), which maps classical input data to quantum states. In this work, we employ an Instantaneous Quantum Polynomial (IQP) encoding scheme to transform a classical data vector $D$ into a quantum state. IQP encoding is a structured, Hamiltonian-inspired strategy that embeds classical information into the phases of a quantum state via commuting gates \cite{bremner2016average, schuld2019quantum, schuld2021effect}.

\begin{equation}
    U_{z}(D) = H^{\otimes n} \, U_{\Phi(D)} \, H^{\otimes n},
\end{equation}
where $n$ is the number of qubits, $H^{\otimes n}$ denotes a layer of Hadamard gates acting on all qubits, and $U_{\Phi(D)}$ is a diagonal unitary that encodes the data-dependent phases. The operator $U_{z}(D)$ is typically applied repeatedly, so that the total depth $r$ corresponds to $r$ repetitions of $U_{z}(D) H^{\otimes n}$.

The core of the IQP encoding is given by
\begin{equation}
    U_{z}(D) = \prod_{(i,j) \in S} R_{YY}\big(\theta_{ij}(D)\big),
\end{equation}

\begin{figure}[!h]
\centering
\includegraphics[width=\linewidth]{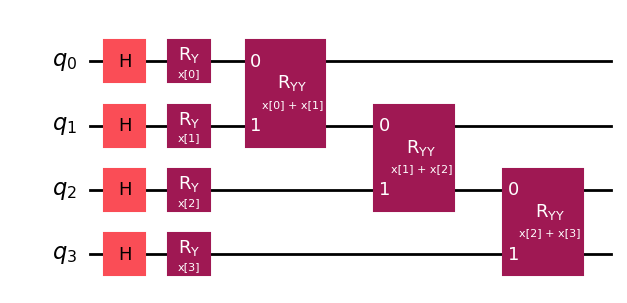}
\caption{IQP encoding }
\label{fig:enter-label}
\end{figure}

where $S$ is a chosen set of qubit pairs, and $R_{ZZ}(\theta_{ij})$ denotes a two-qubit rotation about the $Z \otimes Z$ axis with angle $\theta_{ij}(D)$ that depends on the classical data $D$. The IQP-style encodings may also include single-qubit rotations $R_{X}$, $R_{Y}$, or $R_{Z}$ to enrich the feature map. The set $S$ can be customized to define the entanglement pattern between qubits, thereby controlling the expressivity of the encoded state \cite{bremner2016average}.

The quantum convolutional layer operates analogously to a classical convolutional layer but acts on quantum states instead of classical feature maps. It is composed of several stages: qubit initialization, data encoding (IQP feature map), parameterized unitary operations, and entangling gates.
First qubits are initialized in a superposition state using Hadamard gates $H$ (and optionally rotations such as $R_{x}$), preparing an initial state $|\psi_{\text{in}}\rangle$ (or its density matrix representation $\rho_{\text{in}}$). The data-encoded state is then processed by parameterized unitary blocks,
\begin{equation}
    \rho_{\text{out}} = U(\boldsymbol{\theta}) \, \rho_{\text{in}} \, U^{\dagger}(\boldsymbol{\theta}),
\end{equation}
where $U(\boldsymbol{\theta})$ is a parameterized circuit composed of single-qubit rotation gates $R_x(\theta)$, $R_y(\theta)$, $R_z(\theta)$ and entangling gates such as CNOT. These blocks act as a quantum filter, entangle and rotate subsets of qubits to extract localized features from the IQP-encoded input, in analogy to the receptive fields of classical convolutions.

Following the convolutional operation, a quantum pooling layer reduces the effective dimensionality of the quantum state while retaining the most relevant information. Pooling can be realized through controlled operations, qubit measurements, or partial trace operations that discard selected subsystems. Conceptually, a pooling step can be viewed as mapping a set of density matrices $\{\rho_i\}_{i=1}^{N}$ to a coarser representation $\rho_{\text{out}}$ that aggregates information from multiple qubits or blocks. We employ entangling gates (e.g., CNOT and controlled-SWAP) combined with measurement and partial trace to compress the Hilbert space, analogous to max- or average-pooling in classical CNNs \cite{cong2019quantum, youn2023quantum, maheshwari2025predicting}.

At layer $l$, the QCNN applies a parameterized unitary $U^{(l)}(\boldsymbol{\theta}^{(l)})$ to the input state from the previous layer:
\begin{equation}
    |\psi^{(l)}\rangle = U^{(l)}\big(\boldsymbol{\theta}^{(l)}\big) \, |\psi^{(l-1)}\rangle,
\end{equation}
with pooling operations interleaved between layers to reduce the number of active qubits. The overall quantum circuit depth is kept relatively shallow and scales approximately logarithmically with the number of qubits, $\mathcal{O}(\log n)$, which helps mitigate issues such as vanishing gradients and decoherence on current hardware \cite{cong2019quantum, chen2024resource}.

Training the QCNN involves optimizing the parameters $\boldsymbol{\theta}$ of the convolutional and pooling layers with respect to a task-specific loss function. In our approach, the loss is defined via Hamiltonian-based observable estimation. The effective Hamiltonian used during training is
\begin{equation}
    H = -\sum_{\langle i, j \rangle} J_{ij} Z_{i} Z_{j} - \sum_{i} g_{i} X_{i},
\end{equation}
where $Z_i Z_j$ represents pairwise interactions between qubits $i$ and $j$, and $X_i$ denotes transverse-field terms. To compute observables and gradients, we measure subsets of the quantum state in the Pauli basis, including terms such as $ZIII$, $IZII$, $IIZI$, and $IIIZ$ \cite{peruzzo2014variational, cerezo2021variational}. Expectation values are obtained using an estimator that aggregates measurement statistics across repeated shots. These expectation values are then mapped to the predicted target quantities and used to evaluate the mean squared error (MSE) on the training and testing sets \cite{maheshwari2025predicting}.

After passing through $L$ quantum layers, the final quantum state takes the form
\begin{equation}
    |\psi_{\text{final}}\rangle = U^{(L)}\big(\boldsymbol{\theta}^{(L)}\big) \cdots U^{(1)}\big(\boldsymbol{\theta}^{(1)}\big) |\psi_{\text{in}}\rangle.
\end{equation}
The resulting expectation values are fed into a classical post-processing stage, implemented using PyTorch. This stage consists of fully connected layers (with optional dropout for regularization) that map the quantum features to the four regression targets \cite{liu2025hybrid, chen2024resource, maheshwari2025predicting}.

The QCNN architecture thus follows the quantum circuits to perform feature encoding, convolution, pooling, and Hamiltonian-based observable estimation, while classical neural network layers handle the final regression and regularization. This combination leverages quantum resources such as entanglement and high-dimensional Hilbert spaces for feature extraction, while retaining the flexibility and robustness of classical deep learning for output mapping and optimization.

\section{Result \& Discussion}
All experiments were implemented in Python 3.13.10 using Qiskit v2.2.3 and Qiskit Runtime v0.45. The GWHP dataset comprises \textbf{352 samples in total, of which 264 samples are used for training and 90 for testing}. 4 input features describe each sample, while the QCNN predicts four scalar targets: the X and Y extent of the 1-K isoline, the maximum groundwater temperature, and the position of that maximum within the map. Data normalization was performed using MinMaxScaler and StandardScaler, and the \textbf{batch size was fixed to 16 samples per iteration} for all quantum and classical runs to ensure comparability.

Our work included four sets of experiments and runs on three different backends: a Statevector simulator, a simulated backend with a noise model, and an IBMQ Kyiv QPU backend with and without error mitigation.
The qualitative evaluation of the predicted groundwater temperature fields for the different backends in fig 6. Each panel shows the QCNN’s (or CNN’s) reconstruction of the plume footprint and thermal structure using four output targets: the X and Y extents of the 1-K isoline, the maximum temperature within the plume, and the position of the hottest cell. The corresponding spatial maps reveal how well each model captures the plume geometry and peak temperature.

\begin{figure}[!h] \label{f1}
    \centering
    \includegraphics[width=1.0\linewidth, height = 0.65\linewidth]{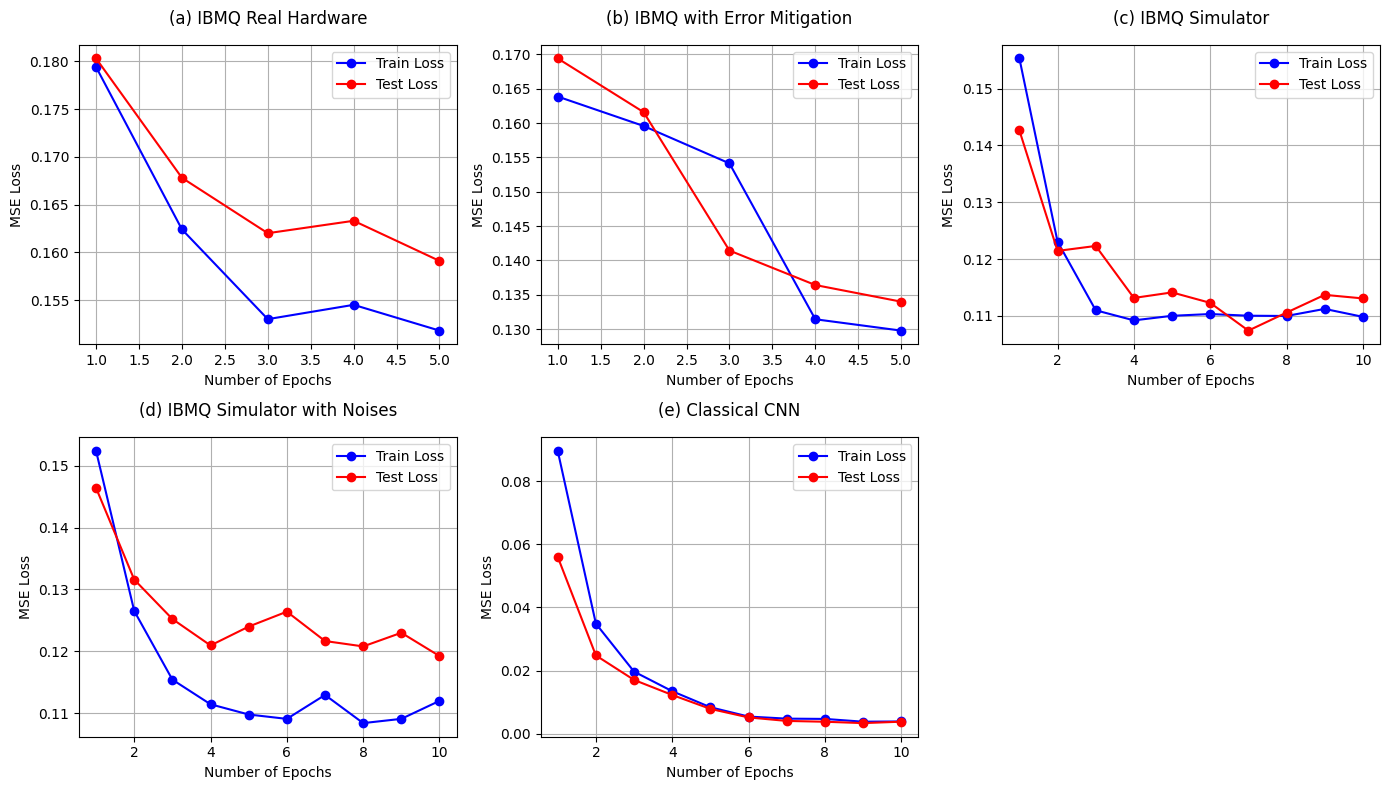}
    \caption{summarizes the training and testing mean squared error (MSE) across five settings: (a) QCNN on the IBMQ Kyiv real hardware, (b) QCNN on Kyiv with advanced error mitigation, (c) QCNN on an ideal statevector simulator, (d) QCNN on a simulator with a realistic noise model, and (e) a classical CNN baseline.}
    \label{fig:qcnn-losses}
\end{figure}

The first quantum experiment was executed on IBM’s 127-qubit Kyiv backend without any dedicated error-mitigation pipeline (Figure~\ref{fig:qcnn-losses}a). The real quantum device deviates considerably from ideal conditions due to gate imperfections, finite coherence times (T\textsubscript{1}/T\textsubscript{2}), readout errors, and crosstalk. These effects are clearly reflected in the \textbf{MSE losses}, which are markedly higher than those obtained from both the ideal and noisy simulators. In this normal regime, the QCNN struggles to consistently reduce the test loss and shows limited generalization capability. This outcome underscores that, for a non-trivial regression task such as predicting spatially resolved thermal characteristics, raw quantum hardware noise remains \textbf{devastating} to QCNN performance.

The QCNN predictions obtained from the IBM Kyiv hardware in fig 6. The color field exhibits noticeable deviations from the expected plume shape, including distortions of the 1-K boundary and inaccuracies in the location and magnitude of the hottest region. These artifacts reflect the high MSE observed in Figure~\ref{fig:qcnn-losses}a and illustrate how hardware noise propagates into the reconstructed spatial pattern.

In the second experiment, we incorporated mitiq \cite{LaRose_2022} error-mitigation techniques on the same Kyiv backend (Figure~\ref{fig:qcnn-losses}b). The mitigation stack included dynamical decoupling sequences (``XX'' and ``XY4'') to reduce dephasing and crosstalk, Twirled Readout Error Extinction (TREX) and measurement mitigation to correct readout errors through randomized measurements and post-processing, and Zero-Noise Extrapolation (ZNE) to extrapolate expectation values back to an effective zero-noise limit. Probabilistic Error Amplification (PEA) and Probabilistic Error Cancellation (PEC) were further employed to calibrate and partially cancel effective noise channels.

Under this mitigation pipeline, both training and testing losses decrease more consistently, and the final MSE values are \textbf{lower than in the normal setting hardware run}. While the mitigated QCNN still does not match the performance of the ideal simulator or the classical CNN, the reduction in error demonstrates that error mitigation can partially recover a useful signal from noisy hardware. This result is consistent with the expectation that, in the absence of fully fault-tolerant devices, sophisticated mitigation schemes are essential for extracting meaningful predictions from quantum circuits.

In contrast, the QCNN predictions when the same circuits are executed on Kyiv with active error mitigation in fig 6(b). Here, the plume outline and peak region appear more consistent with the target pattern: the 1-K isoline is better aligned, and the predicted location of the temperature maximum approaches the expected position. Although residual discrepancies remain, the visual improvement corroborates the reduction in MSE seen in the loss curves. It confirms that error mitigation not only reduces numerical error but also enhances the \textbf{physical interpretability} of the predicted fields.

Across all configurations, the MSE decreases with the number of epochs, indicating that the models can exploit structure in the GWHP data. However, the absolute loss levels differ substantially between backends. As expected, the \textbf{statevector simulator} (Figure~\ref{fig:qcnn-losses}c) achieves the lowest training and testing losses, demonstrating that simultator the QCNN can learn accurate mappings from the four input features to the four target quantities. The \textbf{noisy simulator} (Figure~\ref{fig:qcnn-losses}d) exhibits slightly higher MSE than the ideal simulator but still maintains a stable learning trajectory, suggesting that moderate noise can be tolerated while preserving predictive performance.

The ideal and noisy simulator runs (Figures~6(c) and 6(d)) further highlight this trend. The ideal simulator recovers a plume shape and maximum temperature that closely match the reference distribution, with only minor deviations. The noisy simulator introduces slight blurring and small shifts in the 1-K boundary but largely preserves the overall plume structure, consistent with its intermediate loss values. 

The \textbf{classical CNN} baseline (Figure~\ref{fig:qcnn-losses}e) converges rapidly to very low training and testing losses. This confirms that the underlying regression task is tractable for a well-parameterized classical model and provides a performance lower bound against which the quantum implementations can be compared. In our experiments, the classical model consistently outperforms all quantum variants in terms of final MSE, which is consistent with the current limitations of NISQ-era hardware.

Finally, the classical CNN (Figure~6(e)) provides the visually most accurate reconstruction, capturing both the footprint and peak structure of the plume with high fidelity, which aligns with its lowest MSE among all models.

Overall, the experiments demonstrate both the potential and the current limitations of QCNNs for groundwater heat pump modeling. On ideal simulators, the QCNN can learn the mapping from a compact set of four features to rich spatial descriptors of the plume, achieving competitive performance. However, when deployed on today’s NISQ hardware, strong noise sources lead to substantially degraded performance unless \textbf{aggressive error mitigation} is applied. Even with mitigation, the quantum models remain behind a classical CNN in terms of predictive accuracy.

These findings suggest that, for the GWHP use case studied here, quantum models are not yet ready to replace classical baselines. Instead, they serve as a promising testbed to explore how quantum architectures behave under realistic device conditions and how much performance can be recovered through mitigation. As hardware improves and more efficient mitigation and compilation techniques are developed, QCNNs may become increasingly viable for high-dimensional hydrothermal prediction tasks, especially when integrated into hybrid quantum–classical workflows.

\begin{figure}[!h]
    \centering

    \begin{subfigure}[b]{0.45\textwidth}
        \centering
        \includegraphics[width=\textwidth]{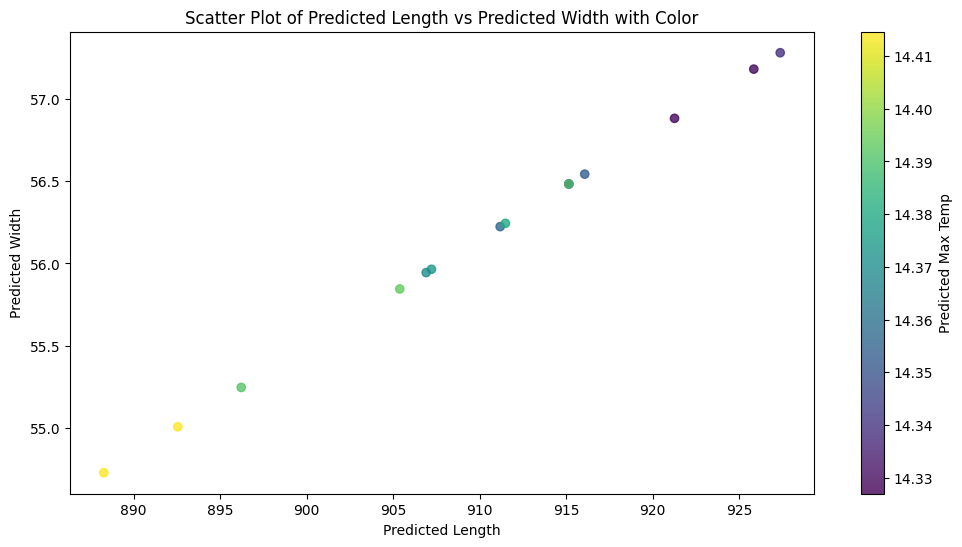}
        \caption{Real Hardware Result}
        \label{fig:img1}
    \end{subfigure}
    \hfill
    \begin{subfigure}[b]{0.45\textwidth}
        \centering
        \includegraphics[width=\textwidth]{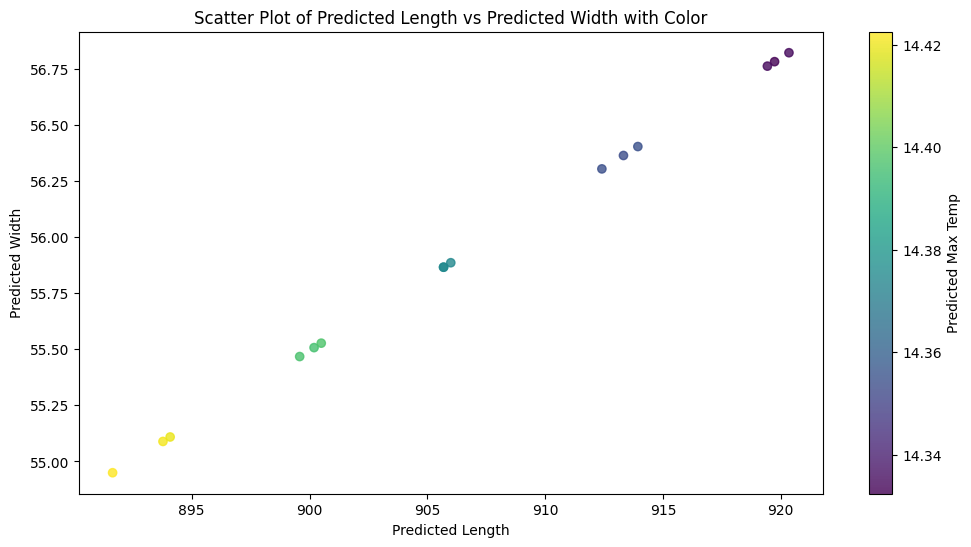}
        \caption{Real Hardware with Error Mitigation}
        \label{fig:img2}
    \end{subfigure}

    \vspace{1em}

    \begin{subfigure}[b]{0.45\textwidth}
        \centering
        \includegraphics[width=\textwidth]{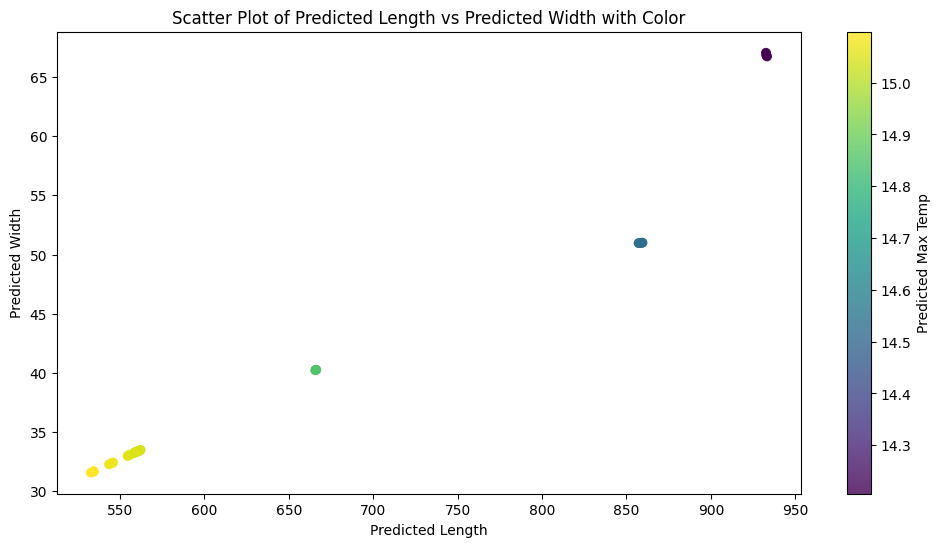}
        \caption{Simulator Result}
        \label{fig:img3}
    \end{subfigure}
    \hfill
    \begin{subfigure}[b]{0.45\textwidth}
        \centering
        \includegraphics[width=\textwidth]{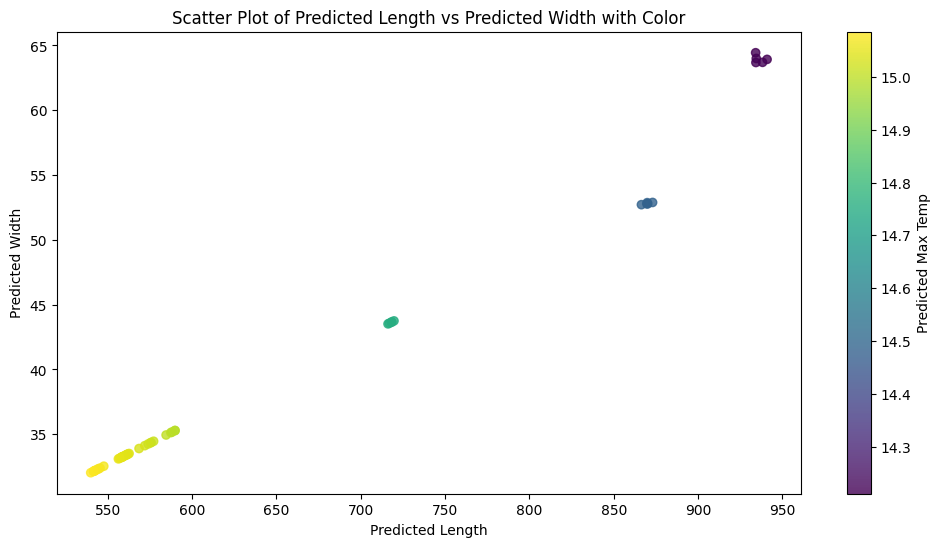}
        \caption{Simulator with Noise}
        \label{fig:img4}
    \end{subfigure}

    \vspace{1em}

    \begin{subfigure}[b]{0.45\textwidth}
        \centering
        \includegraphics[width=\textwidth]{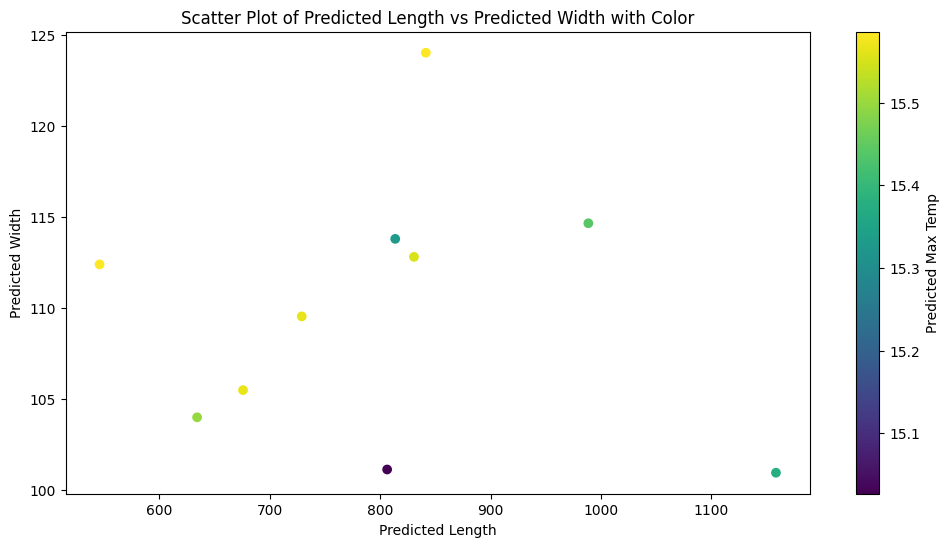}
        \caption{Classical Result}
        \label{fig:img5}
    \end{subfigure}
    \hfill
    \begin{subfigure}[b]{0.45\textwidth}
        \centering
    \end{subfigure}

    \caption{summarizes the Predicated heat plume across five settings: (a) QCNN on the IBMQ Kyiv real hardware, (b) QCNN on Kyiv with advanced error mitigation, (c) QCNN on an ideal statevector simulator, (d) QCNN on a simulator with a realistic noise model, and (e) a classical CNN baseline.}
    \label{fig:all_images}
\end{figure}

\section{Conclusion \& Future Work}

In this work, we rigorously evaluated a Quantum Convolutional Neural Network (QCNN) to predict groundwater heat plume characteristics across four execution regimes. Using 352 samples with standardized preprocessing and a fixed batch size of 16, we benchmarked the QCNN on an ideal statevector simulator, a noisy simulator, and IBM’s 127-qubit Kyiv processor, both without and with advanced error mitigation. The statevector simulator achieved the lowest losses, revealing the QCNN’s full learning capacity under noise-free conditions, whereas realistic noise in simulators and hardware significantly degraded performance. Error-mitigation techniques on Kyiv noticeably improved accuracy and reduced the gap to ideal simulation, but did not eliminate it. Compared to a classical CNN baseline, which remained the most accurate and stable model, the QCNN was competitive only in simulated settings and under mitigated hardware runs. Overall, our results indicate that QCNNs are not yet superior to classical models for this task but show encouraging behavior under mitigation, suggesting growing viability as quantum hardware and noise-control methods advance. Future work will address larger and more diverse plume datasets, scaling to higher-dimensional systems, and cross-platform evaluations on additional quantum processors.

\section*{Acknowledgment}
Funded by Deutsche Forschungsgemeinschaft (DFG, German Research Foundation) under Germany's Excellence Strategy - EXC 2075 – 390740016. We acknowledge the support of the Stuttgart Center for Simulation Science (SimTech) and IBM Quantum for their runtime support.

\bibliography{library}

\end{document}